\documentclass[
superscriptaddress,
 amsmath,amssymb,
 aps,
 twocolumn
]{revtex4-2}

\usepackage[english]{babel}
\usepackage{graphicx} % Required for inserting images
\usepackage{float}
\usepackage{amssymb, amsmath, amsbsy,amsthm}
\usepackage{siunitx}
\usepackage{chemformula}
\usepackage{ulem}
\usepackage{soul}

\begin{document}

% Title, authors and affiliations
\title{Superconducting nitridized-aluminum thin films} 

\author{Alba~Torras-Coloma}
\affiliation{Institut de F\'isica d’Altes Energies (IFAE),
The Barcelona Institute of Science and Technology (BIST), Bellaterra (Barcelona) 08193, Spain}
\affiliation{Departament de F\'isica, Universitat Aut\`onoma de Barcelona, 08193 Bellaterra, Spain}
\author{Leyre~Martínez~de~Olcoz}
\affiliation{Institute of Microelectronics of Barcelona (IMB-CNM), Spanish National Research Council (CSIC), Cerdanyola 08193, Spain}
\author{Eva~C\'espedes}
\affiliation{Institute of Microelectronics of Barcelona (IMB-CNM), Spanish National Research Council (CSIC), Cerdanyola 08193, Spain}
\author{Elia~Bertoldo}
\affiliation{Institut de F\'isica d’Altes Energies (IFAE),
The Barcelona Institute of Science and Technology (BIST), Bellaterra (Barcelona) 08193, Spain}
\author{David~L\'opez-N\'uñez}
\affiliation{Institut de F\'isica d’Altes Energies (IFAE),
The Barcelona Institute of Science and Technology (BIST), Bellaterra (Barcelona) 08193, Spain}
\affiliation{Barcelona Supercomputing Center, Barcelona 08034, Spain}
\affiliation{Departament de F\'isica Qu\`antica i Astrof\'isica and Institut de
Ci\`encies del Cosmos, Universitat de Barcelona, Barcelona 08028, Spain}
\author{Sagar~Paul}
\affiliation{PHI, Karlsruhe Institute of Technology, 76131 Karlsruhe, Germany}
\author{Wolfgang~Wernsdorfer}
\affiliation{PHI, Karlsruhe Institute of Technology, 76131 Karlsruhe, Germany}
\affiliation{IQMT, Karlsruhe Institute of Technology, 76344 Eggenstein-Leopoldshafen, Germany}

\author{Gemma~Rius}\email{gemma.rius@csic.es}
\affiliation{Institute of Microelectronics of Barcelona (IMB-CNM), Spanish National Research Council (CSIC), Cerdanyola 08193, Spain}

\author{P.~Forn-D\'iaz}\email{pforndiaz@ifae.es}
\affiliation{Institut de F\'isica d’Altes Energies (IFAE),
The Barcelona Institute of Science and Technology (BIST), Bellaterra (Barcelona) 08193, Spain}
\affiliation{Qilimanjaro Quantum Tech SL, Barcelona, Spain}

\date{\today}

% ---------------------------- Document starts here ---------------------------

\begin{abstract}
    
    We report the direct observation of superconductivity in nitridized-aluminum thin films. The films are produced by sputtering deposition of aluminum in a controlled mixture of nitrogen diluted in argon. The concentration of applied nitrogen directly determines the properties of the superconducting thin films. We observe samples displaying critical temperatures up to 3.38$\pm$0.01~K and resilience to in-plane magnetic fields well above 1~T, with good reproducibility of the results. This work represents an unambiguous demonstration of tunable superconductivity in aluminum-based nitridized thin films. Our results put forward nitridized aluminum as a promising material to be employed in superconducting quantum circuits for quantum technology applications.

\end{abstract}

\maketitle
\section{Introduction}

Superconducting quantum circuits are a leading platform for building applications in quantum technologies, particularly in the field of quantum computation \cite{Blais2021Circuit, PerezSalinas2021one}. The performance of these circuits is closely tied to the quality of the materials employed, establishing, e.g., limitations on qubit coherence \cite{martinis_decoherence_2005, siddiqi2021engineering} and on resonator quality \cite{Bruno2015reducing, gurevich2023tuning}. One of the leading loss mechanisms is surface dielectric loss in the form of microscopic two-level defects, ubiquitous in both the substrate-metal and metal-air interfaces of the superconducting quantum devices \cite{wang2015surface}. Among the strategies to reduce surface dielectric losses, nitrogen-based superconductors, such as NbN, TiN, or NbTiN, have become of great interest in the quantum computing community of superconducting circuits in the last decade, particularly as materials to build resonators and qubit capacitors \cite{leduc2010titanium, vissers2010low, chang2013improved, bretz2022high, mutsenik2023superconducting, frasca2023nbn}. It is considered that the presence of nitrogen on the surface and within the superconducting thin films helps reducing the impact of two-level system defects that is attributed to native oxides or impurities \cite{DeGraaf2017direct, DeGraaf2018Suppression}, such as those typically found in aluminum and niobium thin films \cite{barends2010reduced}. In addition, a high degree of disorder is typically attributed to very thin ($\lesssim20\,\mathrm{nm}$) nitrogen-based superconductors which explains their characteristic large kinetic inductance \cite{Coumou2013microwave}. Superconducting thin films with high kinetic inductance are key to produce high-impedance circuits with low losses, a very relevant feature required in high-coherence superconducting circuits \cite{Gyenis2021experimental} as well as in single photon detectors \cite{dorenbos2008low, Dorenbos2011low}, state-of-the-art parametric amplifiers \cite{mantegazzini2023high, zhao2023nonlinear}, compact inductance elements and superinductors \cite{niepce2019high}, and kinetic inductance detectors \cite{Levy-bertrand2021subgap, Coiffard2020characterization}.

Motivated by the properties exhibited by nitrogen-based superconductors for quantum circuits, we present here NitrAl (short term for nitridized aluminum), a new superconducting material obtained by depositing aluminum via DC-magnetron sputtering in a controlled atmosphere of nitrogen and argon. To our knowledge, this material has not been analyzed in depth as a superconducting material and only minimal evidence exists on its superconductivity \cite{harris1974superconductivity, morgan1990properties}.

Crystalline aluminum nitride (AlN) is a well-known wide band-gap material with hexagonal crystal lattice commonly used in microelectronics, for example, as a nanomechanical resonator \cite{cleland2001}. In view of its piezoelectric properties it is a suitable candidate for acoustic-wave resonating structures \cite{shu2016characterization}. Recently, AlN has been employed as a bulk acoustic resonator coupled to a superconducting qubit \cite{chou2020}. However, the strong piezoelectricity displayed by AlN results in a loss mechanism through phonon dissipation \cite{cleland2001}. Despite this detrimental effect, thin-film dielectric AlN$_x$ has been used as the junction tunnel barrier in both flux \cite{kim2021enhanced} and transmon qubits \cite{nakamura2011superconducting}, exhibiting long coherence times. The approach followed by both these works consisted on depositing thin films ($\sim$2~nm) in the cubic phase of AlN which does not display piezoelectricity due to its inversion symmetry. On the other hand, tunnel junctions made of AlN have been sputtered for bolometer applications using an inductively coupled plasma source yielding high quality, highly transparent junctions \cite{zijlstra2007epitaxial}. These results indicate that the sources of loss evidenced in wide-bandgap hexagonal AlN can be unimportant in other phases of AlN, existing as e.g. disordered or non-stoichiometric thin films, which may then be used as parts of superconducting circuits.

In contrast to aluminum nitrogen-based superconductors, an oxygen-based superconductor has already been extensively investigated in the form of granular aluminum (grAl) \cite{abeles1966enhancement}. GrAl consists of aluminum grains embedded in an aluminum oxide matrix and can display superconductivity. Actually, this material presents several properties that make it attractive for a variety of applications, such as in quantum computing circuits \cite{grunhaupt2019granular} or employed in astronomy detectors \cite{valenti2019interplay}. GrAl displays a higher critical temperature ($T_{\mathrm{c}}$) than pure Al, up to 3~K for nitrogen-cooled substrates during deposition \cite{cohen1968superconductivity}, and can attain very high room temperature resistivity values, which correlate into superinductor behavior in the superconducting state \cite{grunhaupt2018loss}. 

Our work presents the unambiguous signatures of superconductivity exhibited by a range of NitrAl thin films produced at different concentrations of applied nitrogen (\ch{N2}). We observe the impact of nitrogen concentration on the $T_c$ of the material, as well as the response under large magnetic fields. Our results indicate that NitrAl may be used as a superinductor in quantum circuit applications, as well as in experiments where magnetic field compatibility is desired, such as in semiconductor spin qubits \cite{samkharadze2018strong} and topological-based condensed matter systems \cite{gul2018ballistic}.

\section{Thin Film Materials and Methods}
\label{sec:2}

In this work, we have focused on the dependence of the resulting NitrAl properties as function of the N$_2$ concentration, while keeping the rest of parameters at a constant value. However, other deposition parameters can significantly influence the properties of the films, such as the working pressure or the power value at the target, which will be considered in future studies.

The Al-based thin films are grown by sputtering technique at the Clean Room of the IMB-CNM-CSIC. The equipment is a DC magnetron sputtering from Karl Suss GmbH, model KS-800 HR. The thin film deposition conditions have been developed for substrates at the 4-inches wafer scale. The substrates consist in commercially-available silicon wafers \footnote{The properties of the silicon wafers used for this study are the following. Type/dopant: p/Bor; resistivity: 4-40\SI{}{\ohm\centi\meter}; orientation: $\langle 100 \rangle \pm 1$\SI{}{\degree}; diameter: $100.0\pm0.3$\SI{}{\milli\meter}; thickness: $525\pm 15$\SI{}{\micro\meter}; single sided polished; primary flat: $(101)\pm 1$\SI{}{\degree}, $32.5\pm 2.5$\SI{}{\milli\meter}.} covered with a high quality insulating silicon dioxide layer, typically 500~nm in thickness deposited also in-house. A wet oxidation technique with a processing temperature of 1100~C was applied for obtaining such a silicon dioxide layer. The metal target is supplied as 99.99\% Al, while the processing gases \ch{N2} and \ch{Ar} are of high purity ($>5.5$). The sputtering chamber base pressure is $2\times10^{-7}$~mbar.

The processing parameters for the sputtering deposition, common for all the tests reported in this work, have been established as 60~sccm Ar flow, working pressure $P_w = 5\times10^{-3}$~mbar, and power to the target $P_t = 1000$~W, i.e. those being considered as sputtering conditions within a conventional range. The addition of \ch{N2} is kept in a regime where non-stoichiometric AlN is expected to be obtained, similar to other works using TiN \cite{Vissers2013}. Specifically, the effect of diluted \ch{N2} in the sputtered Al-based thin film properties and functionality has been tested and fully evaluated for \ch{N2}/\ch{Ar} fractional flow mainly in the range of 0\% to 15\%, which correspond to \ch{N2} flow range 0 to 9~sccm (Table~\ref{Table:1}). In this \ch{N2} flow range, the actual deposition rate is determined to be inversely proportional to the applied \ch{N2} flow  (Table~\ref{Table:1}), therefore, the corresponding adjustment of the processing time for each \ch{N2} flow value has been applied. A common nominal or target thickness of $t = 100$~nm for the Al and NitrAl thin films is set. Thin film deposition parameters and methodology reported in this study have been developed and consolidated according to microelectronics standards, to ensure reproducibility in the processing and reliability in the implementation, as well as thin film characteristics, including thickness homogeneity and deposition rate control, and low values of sheet resistance and roughness.

We point out that the actual amount of atomic nitrogen incorporated in the NitrAl thin films produced at variable \ch{N2} flow values could not be reported in this work. Instead, we characterize each nitrogen concentration with the \ch{N2} \% fractional flow value used in the fabrication process.

\begin{table}[b]
\caption{Range of \ch{N2} flow values and the parameters used in the fabrication of NitrAl samples reported. The following parameters are set constant for all samples: Ar flow = 60 sccm, power set to the target $P_t = 1000$~W, nominal thickness deposition $t = 100$~nm, Ar working pressure $P_w = 5\times10^{-3}$~mbar, sputtering chamber base pressure $P_b=2\times10^{-7}$~mbar.\label{Table:1}}
\begin{ruledtabular}
\begin{tabular}{ccccc}
\ch{N2}(sccm) & \ch{N2}/\ch{Ar}(\%) & Rate(nm/s)   \\ \hline
 0.0             & 0.00            & 1.91      \\
 1.5             & 2.50            & 1.78      \\
 2.0             & 3.33            & 1.73      \\
 3.0             & 5.00            & 1.64      \\
 4.0             & 6.67            & 1.55      \\
 5.0             & 8.33            & 1.46      \\
 6.0             & 10.00           & 1.37      \\
 7.0             & 11.67           & 1.28      \\
 8.0             & 13.33           & 1.19      \\
 9.0             & 15.00           & 1.10          
\end{tabular}
\end{ruledtabular}
\end{table}

Once the thin films are produced at wafer scale, each sample is obtained by wafer cleavage in a square geometry with approximate dimensions $\SI{7}{\mm}\times\SI{7}{\mm}$, and fixed to a commercial chip carrier using vacuum grease where it is wire bonded with Al wires. A pair of wire bonds are placed directly on the NitrAl thin films in each of the corners of the sample to improve the electrical contact. Each sample is tested at room temperature by the Van der Pauw technique in this 4-probe configuration. The resistance of the film is obtained by fitting the resulting current-voltage (IV) curve measured with a source-meter instrument. To calculate the sheet resistance, we multiply the resistance obtained from the fit by the Van der Pauw constant: $\pi / \ln{2}$ \cite{van1958method}.

After the room temperature validation, each sample is mounted either on the mixing chamber stage or on the still plate of a dry dilution refrigerator with a base temperature of 20~mK. Standard magnetic shielding is employed both at room temperature and at low temperature consisting in high-$\mu$ materials. Accurate temperature control is implemented in the determination of $T_{\mathrm{c}}$ values of the thin films. As the $T_{\mathrm{c}}$ of most samples lies in the 1-3.5~K range, specific procedures to stabilize the temperature in this range are required (see Sec.~A of the Supplementary Material for details). 

In order to capture the entire evolution of resistance with temperature, $R(T)$, the resistance of each thin film is monitored as the system cools down/warms up. The range of swept currents is gradually adjusted to avoid excessive heat dissipation when the samples are in the normal state. Once in the superconducting state, the current is adjusted to determine the switching current value at base temperature. Then, the current range is reduced significantly below the switching current value in order to perform the $T_{\mathrm{c}}$ measurements. 

Magnetic field-dependent IV-characteristics measurements of NitrAl samples are performed in a different dilution refrigerator in the KIT laboratory equipped with a vector magnet, as described in \cite{thirion2002micro}. Using this experimental setup, perpendicular fields $B_{\mathrm{z}}$ up to 0.1~T and in-plane fields $B_{\mathrm{x}}$ up to 1.4~T
can be applied without manually rotating the sample, while the bath temperature can be varied down to 30~mK. In this case, the approximate dimensions of the samples are reduced to $\SI{1}{\mm}\times\SI{3}{\mm}$ to fit the special chip carrier used. The resistance values are obtained from the IV-characteristics measured in lock-in configuration about zero constant bias current $I_b$ and $\SI{10}{\micro\ampere}$ modulation amplitude, as a function of the temperature and the magnetic field applied in the different directions. More details can be found in Sec.~D1 of the Supplementary Material.

\section{Functional characteristics of the NitrAl thin films}
\label{sec:3}

\subsection{Electrical properties at room temperature}

\begin{figure}
         {\includegraphics[]{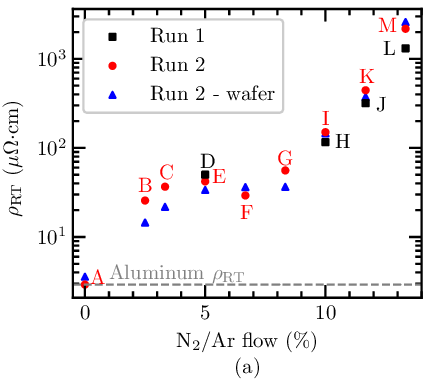}}
\quad
         {\includegraphics[]{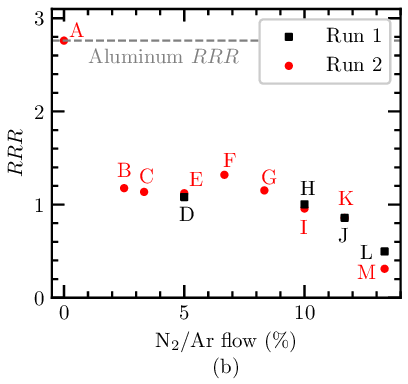}}
        \caption{(a) Resistivity $\rho_{\mathrm{RT}}$ measured for each NitrAl film at room temperature as function of \ch{N2} flow, relative to the fixed \ch{Ar} flow used in the sputtering process. The samples labeled in black (red) were produced in the first (second) run of fabrication. The blue triangles correspond to the averaged values over a wafer scale. The Al reference values are marked with a gray dashed line for reference. (b) $RRR \equiv \rho_{\mathrm{RT}}/\rho_{\mathrm{4K}}$ as function of \ch{N2} flow, $\rho_{\mathrm{4K}}$ being the resistivity at \SI{4}{K}.\label{fig:resistivity_vs_N2}}
\end{figure}

In Fig.~\ref{fig:resistivity_vs_N2}(a) we show the resistivity values at room temperature $\rho_{\mathrm{RT}}$ for the range of thin films produced, plotted as a function of the ratio of \ch{N2}/\ch{Ar} flow in the process of sputtering deposition, and for two different fabrication runs (red and black markers). The blue triangle markers represent direct determination of $\rho_{\mathrm{RT}}$ averaged over several points randomly chosen across the 4" wafer scale, for which dispersion is found to be about or below 3-5\%. The resistivity estimations at the sample level based on the Van der Pauw technique at chip level agree with the wafer-scale averaged values. 

The obtained data show that the addition of \ch{N2} flow in the sputtering process clearly impacts the thin film resistivities for all flow values, with a significant increase with respect to the 0\% reference thin film (pure Al) whose value is close or nearly comparable to bulk Al. The results for the two different fabrication runs of sputtered NitrAl thin films show significant reproducibility. $\rho_{\mathrm{RT}}$ steadily rises up to an order of magnitude with respect to pure Al in the range 2.5\%--5\%. Then, between 5\% and 8.33\% $\rho_{\mathrm{RT}}$ flattens. Beyond 10\%,$\rho_{\mathrm{RT}}$ continues to rise, following a divergent dependence, finally displaying an electrically insulating characteristic at 15\% \ch{N2} flow, with $\rho_{\mathrm{RT}}\sim1.5~\Omega\,$cm, even visible in the surface color of the sample (see Supplementary Material E for more details).

\subsection{Electrical properties at low temperatures}

In Fig.~\ref{fig:resistivity_vs_N2}(b), we display the residual resistivity ratio, $RRR \equiv \rho_{\mathrm{RT}}/\rho_{\mathrm{4K}}$ with $\rho_{\mathrm{4K}}$ the resistiviy measured at \SI{4}{K}, as function of \ch{N2} flow down to 4~K. The decrease in $RRR$ as \ch{N2} is incorporated is a clear indicator of increased disorder and impurities in the thin film \cite{Ficket1971Aluminum}. Between 2.5\% up to 10\%, $RRR\gtrsim1$. Above 10\%, $RRR$ tends to decrease below 1. This behaviour is reminiscent of grAl devices \cite{levy2019electrodynamics}, and it is in contrast with, e.g., NbN disordered films, where the conductivity of the material gradually decreases as disorder increases \cite{chand2012phase}. Exemplary IV curves for several samples at different temperatures are shown in Fig.~\ref{fig:IVs} of the Supplementary Material. 

An abrupt increase of resistivity at higher \ch{N2} flows [Fig.~\ref{fig:resistivity_vs_N2}(a)] is compatible with a Superconductor-to-Insulator Transition (SIT), either driven by localization of conduction electrons (Mott) or by disorder (Anderson) \cite{Gantmakher2010superconductor}. The values of $\rho_{\mathrm{RT}}$ displayed by the NitrAl samples investigated are in the same order of magnitude as grAl devices previously reporting a Mott insulator-like phase transition \cite{levy2019electrodynamics, grunhaupt2018loss}, as well as disordered NbN and NbTiN thin films displaying an Anderson-like phase transition \cite{chand2012phase, Driessen2012strongly}. 

\begin{figure}[!hbt]
    \includegraphics[]{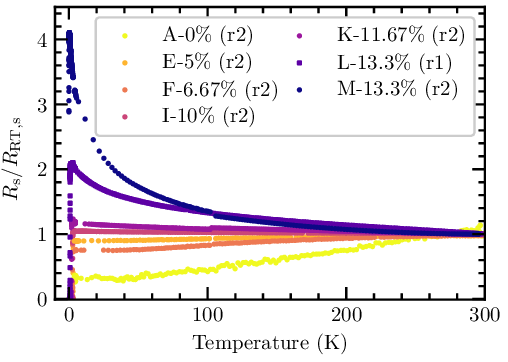}
    \caption{Normalized sheet resistance curves for several NitrAl samples. The samples with \ch{N2} flow in the range 0\% to 6.67\% exhibit metallic-like behavior with $RRR>$1. By contrast, the samples with \ch{N2} flow in the range 11.67\%-13\% exhibit insulator-like behavior with $RRR<1$. In fact, the sample 13.3\% (r2) remains resistive down to the base temperature of our refrigerator (19~mK). The transition point between the two regimes appears around the sample with 10\% \ch{N2} flow with $RRR\sim1$.}
    \label{fig:RvsT_plots}
\end{figure}

 Figure~\ref{fig:RvsT_plots} shows the normalized sheet resistance curves $R_s/R_{RT, s}$ obtained for different NitrAl samples in the range between \SI{19}{\milli\kelvin} and 300~K. As the \ch{N2} flow is increased, we observe two different regimes which we classify according to the change in slope with temperature on the different curves, as follows: 

\begin{itemize}
    \item \textit{Metallic-like} $(\mathrm{d} R_{\mathrm{s}} / \mathrm{d} T > 0)$: In this regime the resistance monotonically decreases as temperature decreases ($RRR>1$). For the studied set of NitrAl films, the metallic regime comprises samples produced with \ch{N2} flows between $0\%$ and $8.33\%$ or, equivalently, for samples with $\rho_{\mathrm{RT}}\lesssim\SI{76}{\micro\ohm\,\centi\meter}$.
    
    \item \textit{Transition point} $(\mathrm{d}  R_{\mathrm{s}} / \mathrm{d} T \sim 0)$: it is observed around the sample processed with $10\%$ \ch{N2} flow, which corresponds to $\rho_{\mathrm{RT}}\sim115$--$\SI{151}{\micro\ohm\,\centi\meter}$ $(RRR\sim1)$. 
    
    \item \textit{Insulator-like} $(\mathrm{d} R_{\mathrm{s}} / \mathrm{d}T < 0)$: this regime comprises samples with increasing resistance as temperature decreases ($RRR<1$). We include in this category samples produced with $11.67\%$ and $13.3\%$ of \ch{N2} flow, or equivalently, with $\rho_{\mathrm{RT}}\gtrsim\SI{396}{\micro\ohm\,\centi\meter}$. For this group of samples, we can also spot two distinctive features. For films produced with $11.67\%$ of \ch{N2} fractional flow, the increase in resistance is quasi-linear for decreasing temperatures. On the other hand, the most resistive samples ($13.3\%$) show an accelerated, exponential-like increase of resistance as temperature is lowered.  
\end{itemize}

\begin{figure*}
     \includegraphics[]{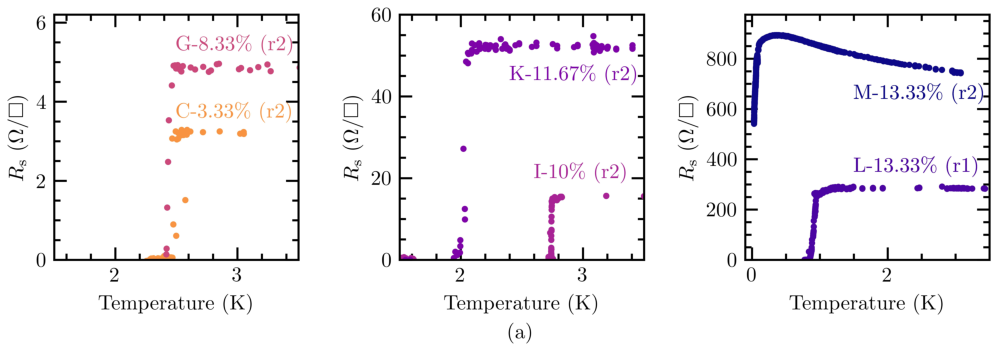}
     \quad
         \includegraphics[]{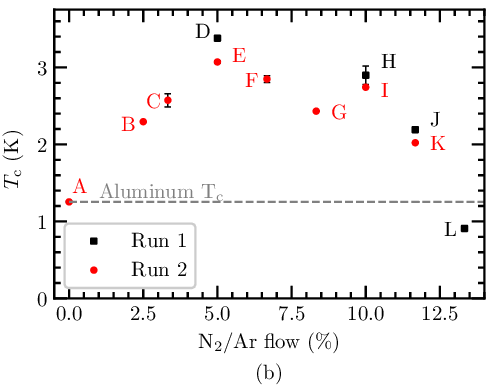}
     \quad
         \includegraphics[]{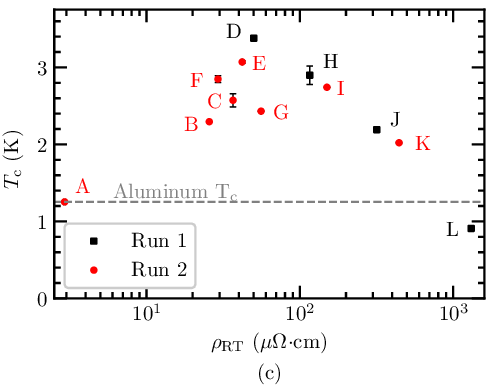}
        \caption{(a) Sheet resistance measurement near the normal metal-to-superconductor transition for different samples. Note the decrease in slope at the transition temperature for increasing \ch{N2} flow. (b)-(c) Critical temperatures of all NitrAl samples determined by resistivity measurements as function of (b) the fraction of \ch{N2} flow during fabrication and (c) the resistivity measured at room temperature [Fig.~\ref{fig:resistivity_vs_N2}(a)]. Samples produced in the first (second) run of fabrication are shown in black squares (red circles). The $T_{\mathrm{c}}$ of aluminum (0\% reference sample) is shown with a gray dashed line for reference.}
        \label{fig:Tc_plots}
\end{figure*}

Below 4~K, all NitrAl samples in the 2.5\%-13.3\% \ch{N2} flow range show a normal-to-superconducting state transition at varying $T_{\mathrm{c}}$ values. Figure~\ref{fig:Tc_plots}(a) shows the suppression of resistance at the transition temperature $T_{\mathrm{c}}$ for samples with different \ch{N2} flow fractions. We take the definition of $T_{\mathrm{c}}$ as the point where the resistance drops $50\%$ with respect to the onset point of $R(T)$ \cite{poole2014superconductivity}.  We did not observe the superconducting transition above \SI{19}{\milli\kelvin} for sample $13.33\%$ produced in the second run of fabrication (r2), indicating that the SIT may lie close to that process \ch{N2} flow value. 

In several samples we have observed an increase of 20\%-30\% of the resistance prior to the transition to the superconducting state. Such a peak appears only for particular combination of current and voltage probes. Annex~\ref{sec:ann_b_2} gives more details about this feature, which we attribute to the disordered nature of the thin films.

Figures~\ref{fig:Tc_plots}(b)-(c) summarize the obtained critical temperatures as a function of \ch{N2} flow and $\rho_{\mathrm{RT}}$, respectively. Overall, the trend of $T_{\mathrm{c}}$ values changes non-monotonically, with most samples exceeding the $T_{\mathrm{c}}$ of pure Al (dashed gray-line). The highest $T_{\mathrm{c}} = 3.38\pm\SI{0.01}{\kelvin}$ is observed at 5\% \ch{N2} flow \footnote{It is likely that and even higher $T_{\mathrm{c}}$ than the one observed at 5\% lies in a \ch{N2} flow not sampled by the discrete set of values used in this work.}, while the 13.33\% \ch{N2} flow sample lies below Al with $T_{\mathrm{c}} = 0.86\pm\SI{0.02}{\kelvin}$. The general shape of the $T_{\mathrm{c}}$ curve is dome-like, both in percentage and resistivity $\rho_{\mathrm{RT}}$. However, the samples with $6.67\%$ and $8.33\%$ \ch{N2} flow do not seem to evenly follow the trend of a dome. On one hand, they display a drop in $T_{\mathrm{c}}$ in the center of the dome, and, on the other hand, their values of $\rho_{\mathrm{RT}}$ versus $T_{\mathrm{c}}$ in Fig.~\ref{fig:Tc_plots}(c) do not follow the rest of samples. A more in-depth study of NitrAl materials at the microscopic level would be required to interpret these particular features.

We note that the dome-shape of $T_{\mathrm{c}}$ versus $\%\ch{N2}$ in Fig.~\ref{fig:Tc_plots}(b) roughly defines three regions: increasing $T_{\mathrm{c}}$ (0\%-5\%), saturation of $T_{\mathrm{c}}$ (5\%-10\%) and decreasing $T_{\mathrm{c}}$ (10\%-13.33\%), which approximately correspond with the three regimes described in Fig.~\ref{fig:resistivity_vs_N2}(a) of $\rho_{\mathrm{RT}}$ versus \%\ch{N2}/\ch{Ar}. This correspondence points to a possible connection between $\rho_{\mathrm{RT}}$ and $T_{\mathrm{c}}$ driven by the same mechanism, likely the disorder and/or localization of the superconducting wave function \cite{Gantmakher2010superconductor}. Indeed, from the data displayed in Fig.~\ref{fig:RvsT_plots}, the insulator-like behavior begins at the 10\% samples. 

Three different regimes of $T_{\mathrm{c}}$ versus $\rho$ have also been observed in other disordered superconductors such as grAl \cite{dynes1981metal}. In the case of grAl it is argued that a competitive effect occurs with increasing concentration of oxygen between the phase stiffness $J$ of the superconducting condensate, the charging energy $E_{\mathrm{C}}$ for a single electron to tunnel between aluminum grains, and the superconducting gap $\Delta$ \cite{levy2019electrodynamics}. In the metallic regime, a single wave function is formed by all electrons which is characterized by a phase stiffness $J$, orders of magnitude above $E_{\mathrm{C}}$ and $\Delta$ [that would correspond to $\rho_{\mathrm{RT}}$ values below $\sim\SI{40}{\micro\ohm\,\centi\meter}$ for our NitrAl thin films, Fig.~\ref{fig:Tc_plots}(c)]. At the dome maximum, the phase stiffness $J$ falls below the Coulomb repulsion [$\rho\sim\SI{50}{\micro\ohm\,\centi\meter}$ in Fig.~\ref{fig:Tc_plots}(c)]. At this point, the transition towards an insulator begins \cite{Gantmakher2010superconductor}. When $J\sim\Delta$, the Coulomb repulsion localizes the wave function in the grains and the material arrives at the SIT [$\rho>\SI{1000}{\micro\ohm\,\centi\meter}$ in Fig.~\ref{fig:Tc_plots}(c)].

This qualitative description of grAl appears to fit well to our observations with NitrAl despite that we have no evidence of the latter having a granular structure. Yet, there exist some differences. Generally, in disordered and granular superconductors the SIT is expected for resistances close to the quantum of resistance $R_{\mathrm{Q}} = h/(2e)^2 \approx 6.4~\text{k}\Omega$. For the present study, the $R_{\mathrm{4K,s}}$ threshold value for the suppression of superconductivity is $700.8 \pm 0.9$~$\Omega / \square$, which is almost an order of magnitude smaller than $R_{\mathrm{Q}}$. We note that the observations of Ref.~\cite{levy2019electrodynamics} are made on thinner films (20~nm thick), which likely enhance the effects by increasing the film resistance in the normal state. For instance, we do not have evidence of features appearing at $\sim1.2$~K reminiscent of pure Al for the insulating films, which might derive from the core-shell structure of grAl. The thickness-dependent effects and the microscopic material description of NitrAl will be the subject of future works, where these features will be explored explicitly in more detail. 

We are in the process of building a microscopic model of NitrAl, necessary to draw quantitative conclusions from the data in Fig.~\ref{fig:Tc_plots} to understand the dominant energy scales in each range of $\rho_{\mathrm{RT}}$, and to determine the type of SIT. 

The values of $T_{\mathrm{c}}$ and $\rho_{\mathrm{4K}}$ can be used to estimate the sheet kinetic inductance ($L_{\mathrm{k,s}}$) of the films by applying the Mattis-Bardeen formula for the complex conductivity in the local, dirty limit at low frequency $(hf\ll k_{\mathrm{B}} T)$ and in the low temperature limit $(T \ll T_{\mathrm{c}})$ \cite{Rotzinger2017aluminum}, 
\begin{equation}
    L_{\mathrm{k,s}} = 0.18 \frac{\hbar R_{\mathrm{4K,s}}}{k_B T_{\mathrm{c}}},
    \label{eq:Kinetic_inductance}
\end{equation}
where $R_{\mathrm{4K,s}}$ is the normal state sheet resistance at \SI{4}{\kelvin}, $\hbar$ is the reduced Planck constant and $k_{\mathrm{B}}$ stands for the Boltzmann constant. The values of sheet kinetic inductance obtained for the NitrAl samples range between $1.31\pm 0.01$ pH/$\square$ (0\% N$_2$) and $422.48 \pm 8.22$ pH/$\square$ (13.3\% N$_2$). The complete set of  values for $L_{\mathrm{k,s}}$, together with $T_{\mathrm{c}}$, $\rho_{\mathrm{RT}}$ and other sample parameters, are shown in Table~\ref{tab:results}. This preliminary estimation of $L_{\mathrm{k,s}}$ points towards NitrAl as a material with high$L_{\mathrm{k,s}}$, especially for the most resistive samples (11.67\%-13.33\% \ch{N2} flow) that reach values close to other reported superinductive materials such as grAl~\cite{maleeva2018circuit}. 
\begin{table*}
\caption{Observed parameters for the different NitrAl samples measured: sample label, \ch{N2}/\ch{Ar} fractional flow (\%), fabrication run number, nominal thickness $t$, critical temperature $T_{\mathrm{c}}$, resistivity at room temperature $\rho_{\mathrm{RT}}$, and at \SI{4}{\kelvin} $\rho_{\mathrm{4K}}$, estimated sheet kinetic inductance $L_{\mathrm{k,s}}$ using Eq.~(\ref{eq:Kinetic_inductance}).\label{tab:results} }
%\centering
\begin{ruledtabular}
\begin{tabular}{cccccccc}
Sample & \ch{N2}/\ch{Ar} (\%)   &   Run & $t$ (\SI{}{\nano\meter}) &  $T_{\mathrm{c}}$(\SI{}{\kelvin})  & $\rho_{\mathrm{RT}}$ (\SI{}{\micro\ohm \centi\meter}) & $\rho_{\mathrm{4K}}$ (\SI{}{\micro\ohm \centi\meter}) &   $L_{\mathrm{k,s}}$ (pH/$\square$) \\ \hline
A & $0$ - Al ref.      &   2   &     $100.0$             &    $1.25\pm 0.01$      & $2.92 \pm 0.04$  &  $ 1.05 \pm 0.05 $    &          $0.12 \pm 0.01$ \\
B & $2.50$             &   2   &     $100.0$             &    $2.30\pm 0.01$      & $25.66 \pm 0.03$ &  $ 21.82 \pm 0.03 $   &          $1.31 \pm 0.01$ \\
C & $3.33$             &   2   &     $100.0$             &    $2.57\pm 0.08$      & $36.67 \pm 0.03$ &  $ 32.28 \pm 0.11 $   &          $1.72  \pm 0.06$  \\
D & $5.00$             &   1   &     $100.0$             &    $3.38\pm 0.01$      & $50.01 \pm 0.03$ &  $ 46.27 \pm 0.03$    &          $1.88  \pm 0.01$  \\
E & $5.00$             &   2   &     $100.0$             &    $3.07\pm 0.02$      & $42.07 \pm 0.13$ &  $ 37.50 \pm 0.04 $   &          $1.68 \pm 0.01$ \\
F & $6.67$             &   2   &     $100.0$             &    $2.85\pm 0.04$      & $29.25 \pm 0.03$ &  $ 22.17 \pm 0.03 $   &          $1.07  \pm 0.02$  \\
G & $8.33$             &   2   &     $100.0$             &    $2.43\pm 0.01$      & $55.88 \pm 0.03$ &  $ 48.50 \pm 0.03 $   &         $2.74   \pm 0.01$   \\
H & $10.0$             &   1   &     $83.0$              &    $2.89\pm 0.12$      & $115.97\pm 0.16$ &  $ 115.63 \pm 0.03 $  &          $6.61  \pm 0.27$  \\
I & $10.0$             &   2   &     $100.0$             &    $2.74\pm 0.01$      & $150.26\pm 0.03$ &  $ 156.91 \pm 0.03 $  &          $7.88  \pm 0.01 $   \\
J & $11.67$            &   1   &     $97.1$              &    $2.19\pm 0.01$      & $317.52\pm 0.03$ & $ 370.11 \pm 0.03 $  &         $23.91 \pm 0.08$           \\
K & $11.67$            &   2   &     $100.0$             &    $2.02\pm 0.01$      & $444.13 \pm 0.03$&  $ 458.59 \pm 0.03 $  &          $35.30 \pm 0.10$    \\
L & $13.33$            &   1   &     $94.6$              &    $0.91\pm 0.02$      & $1312.25 \pm 0.03$  &  $ 2641.66 \pm 0.05 $ &       $422.48 \pm 8.22$   \\
M & $13.33$            &   2   &     $100.0$             &      -                 & $2177.01\pm 0.03$&  $ 7008.47 \pm 0.03 $   &              -             \\
\hline
\end{tabular}
\end{ruledtabular}
\end{table*}

\begin{figure}[!hbt]
         {\includegraphics[]{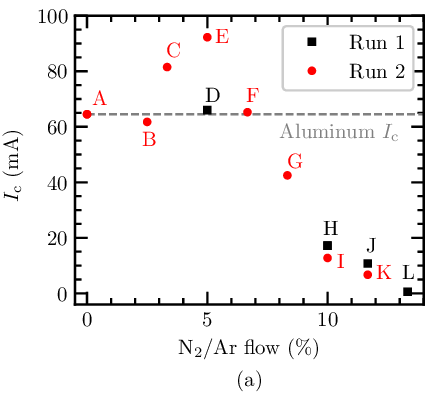}}
         \label{fig:Ic_percs}
     \quad
         \centering
         {\includegraphics[]{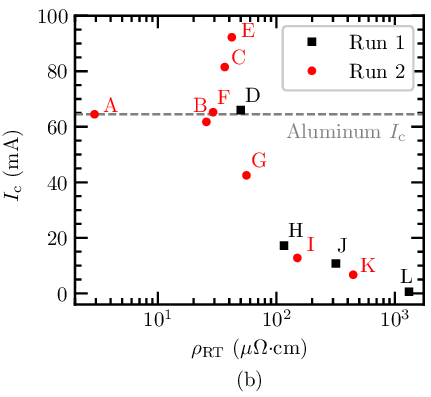}}
         \label{fig:Ic_resist}
        \caption{Critical currents $I_{\mathrm{c}}$ of NitrAl as a function of (a) the fraction of \ch{N2} flow during fabrication and (b) the resistivity $\rho_{\mathrm{RT}}$ at room temperature of the samples.}
        \label{fig:Ic_plots}
\end{figure}

Additionally, the critical currents $I_{\mathrm{c}}$ have been measured at base temperature for each \ch{N2} flow value. Since these measurements were performed on thin films with no device pattern, the results only provide a qualitative understanding of the behavior of $I_{\mathrm{c}}$ in NitrAl films. The resulting values are shown in Fig.~\ref{fig:Ic_plots}. Note that when the samples from both fabrication runs are sorted by $\rho_{\mathrm{RT}}$ (Fig.~4b) they follow an overall consistent trend, which is in contrast to the deviations between the $I_{\mathrm{c}}$ from different fabrication runs for a given N$_2$ flow value (Fig.~4a). For samples in the range of 8.33\%-13.33\% the critical current lies below that of Al, possibly indicating that the superconducting state is weakening due to the increase in disorder/localization. 

\subsection{Magnetic field response}

\begin{figure}[!htb]
    \includegraphics[]{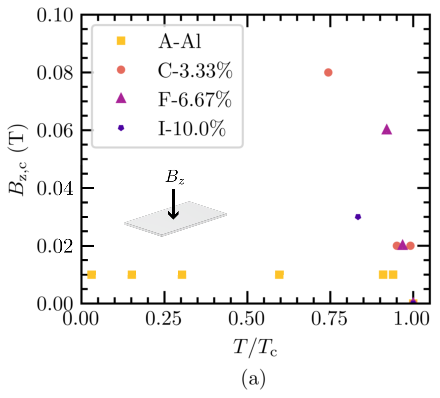}
    \quad
    \includegraphics[]{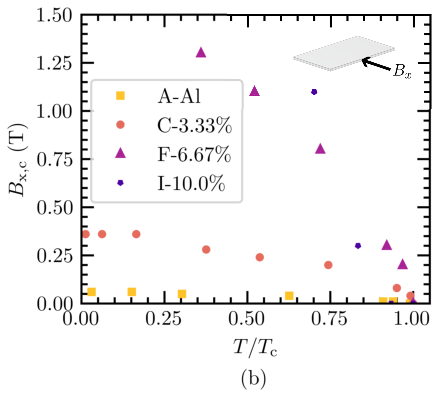}     
    \caption{Magnetic fields at which superconductivity is destroyed as function of the normalized temperature in the (a) $z$-direction, $B_{\mathrm{z,c}}$ (perpendicular to the chip) and (b) the $x$ direction, $B_{\mathrm{x,c}}$ (parallel to the chip) for different NitrAl samples produced in the second run of fabrication. The temperatures have been normalized to the critical temperature of each sample. In (a), the reference Al sample shows no temperature dependence as the smallest applicable field of 0.01~T already drives it in the normal state at base temperature.}
    \label{fig:B_maps}
\end{figure}

Figure~\ref{fig:B_maps} shows the extracted fields at which superconductivity is destroyed $B_{\mathrm{c}}$ for perpendicular $B_{\mathrm{z,c}}$ (a) and in-plane $B_{\mathrm{x,c}}$ (b) externally applied magnetic fields as function of temperature.   

At base temperature, the Al reference ($0\%$ \ch{N2} flow) has a critical field below \SI{0.01}{\tesla} for $B_{\mathrm{z}}$ and \SI{0.06}{\tesla} for $B_{\mathrm{x}}$, as expected for thin-film Al \cite{tinkham2004introduction}. For the subset of NitrAl samples displayed in Fig.~\ref{fig:B_maps}(a), the available range of the magnet in  $B_{\mathrm{z}}$ is only capable of driving the system out of the superconducting state close to the critical temperature of the NitrAl samples. On the other hand, for $B_{\mathrm{x}}$ [Fig.~\ref{fig:B_maps}(b)] the increase of resilience of NitrAl to magnetic fields with respect to the Al reference sample is high but less pronounced. We observe an increase of almost a factor 7 in the $B_{\mathrm{x, c}}$ for sample $3.3\%$. For the $10\%$ sample, the available range of the magnet in$B_{\mathrm{x}}$ is not enough to destroy the superconducting state at base temperature. 

Measurements of IV-characteristics as function of the applied magnetic fields and temperature can be found in Fig.~\ref{fig:IVs_B} of the Supplementary Material. In particular, for the 10\% sample a gradual increase of resistance is observed for the largest applied fields close to the switching current of the sample. This effect is reminiscent of type-II superconductivity where such a resistance at large applied fields close to the switching current of the sample is due to vortex flow induced by the applied current~\cite{tinkham2004introduction}. In light of this connection with type-II superconductors, the observed field at which superconductivity is destroyed may be identified as $B_{\mathrm{c}} = \mu_0H_{\mathrm{c2}}$. Additional measurements under magnetic fields are needed to study the type of superconductivity displayed by NitrAl, but those are beyond the scope of this initial work.

The results obtained for NitrAl as function of the magnetic field follow the observations in other disordered superconductors \cite{Bachar2015Mott, Abeles1967Critical}, where an increase in resistivity and superconducting gap with respect to the pure superconductor is accompanied by an increase in critical field. 

Overall, there is a clear increase in the field at which superconductivity is destroyed in both axes  $B_{\mathrm{z}}, B_{\mathrm{x}} $ for increasing $\rho_{\mathrm{RT}}$ in the NitrAl samples. This magnetic field resilience makes NitrAl an interesting candidate to be used in circuits where high magnetic fields are required, such as, e.g., those manipulating electron spins \cite{samkharadze2018strong} or hybrid superconductor-semiconductor structures \cite{kringhoj2021magnetic}. The behaviour of NitrAl thin films under strong magnetic fields is comparable to that displayed in general by disordered superconductors, including grAl \cite{borisov2020high}, NbN \cite{chand2012phase}, and NbTiN \cite{muller2022magnetic}.

\section{Conclusions}
\label{sec:4}

We have reported in this work an in-depth characterization of the superconducting properties of nitridized aluminum -NitrAl-, a new superconducting material that we put forward to be introduced in quantum circuits for quantum technology-related applications. By adjusting the relative concentration of nitrogen with respect to argon in the sputtering process to form the NitrAl films, we are able to control its superconducting and room temperature characteristics. For a wide parameter range, NitrAl samples display properties superior to pure Al, such as higher critical temperature (up to 3.38$\pm$0.01~K) and larger fields at which superconductivity is destroyed (above 1~T for in-plane fields). The controllability of the fabrication process, the range of reported critical temperatures and the extent of measured resistivities open the possibility to use NitrAl as a superinductor material, as well as a superconductor with adjustable $T_{\mathrm{c}}$.

Further studies are required to infer the microscopic properties of NitrAl, particularly to discern whether the likely presence of native oxide is responsible for some of the properties observed. In addition, its behavior needs to be investigated as a material for high-frequency circuits in the form of resonators or even superconducting qubits, particularly the losses exhibited by NitrAl under these conditions. 

\section*{Data availability statement}
Data available upon request or in the following link: \url{https://doi.org/10.5281/zenodo.10194875}.

\section*{Acknowledgements}
The authors would like to thank Karl Jacobs from I. Physikalisches Institut, Universitaet zu Koeln, for providing the first set of NitrAl samples and for insightful discussions. We would also like to acknowledge Fabio Henriques for helping with the temperature measurements, Juan Jim\'enez for developing software for data analysis, and fruitful discussions with E.~Driessen, T.~Klapwijk, and A.~Endo. We acknowledge funding from the Ministry of Economy and Competitiveness and Agencia Estatal de Investigaci\'on (RYC2019-028482-I, RYC-2016-21412, FJC2021-046443-I, PCI2019-111838-2, PID2021-122140NB-C31, PID2021-122140NB-C32, TED2021-130292B-C44), the European Commission (FET-Open AVaQus GA 899561, QuantERA), and program ‘Doctorat Industrial’ of the Agency for Management of University and Research Grants (2020 DI 41; 2020 DI 42). IFAE is partially funded by the CERCA program of the Generalitat de Catalunya. This study was supported by MICIIN with funding from European Union NextGenerationEU(PRTR-C17.I1) and by Generalitat de Catalunya.

\bibliographystyle{iopart-num}
\bibliography{bibliography.bib}

\newpage

\renewcommand{\thesubsection}{\Alph{subsection}}

\onecolumngrid
\section*{Supplementary Material}

\subsection{IV current caracteristics}
\label{sec:S-A}
The dilution refrigerator is in a stable configuration with the pulse tube ON while the 4K, Still, and mixing chamber (MXC) plates are thermally connected. No 3He/4He mixture is circulating.
A small quantity (20-50 mbar) of mixture (total $\sim500$~mbar) is introduced in the system. When all plates are at the same temperature, the MXC and Still plates are thermally isolated, while the Still remains in thermal contact with the 4K plate. In this configuration, the MXC can be stabilized using its heater to temperatures below the ones reachable by only employing the pulse tube (3-4K). If the Still is then isolated from the 4K plate and more mixture is condensed in a controlled way, the system can slowly reach temperatures $\sim1$~K or lower. The temperature of the mixing chamber and the Still can be modified by using their individual heaters and/or by restoring the thermal connection between the two plates. The whole procedure can be sped up or slowed down by controlling the amount of mixture in circulation and the heat introduced by the Still and MXC heaters.

\begin{figure}[H]
         \includegraphics[]{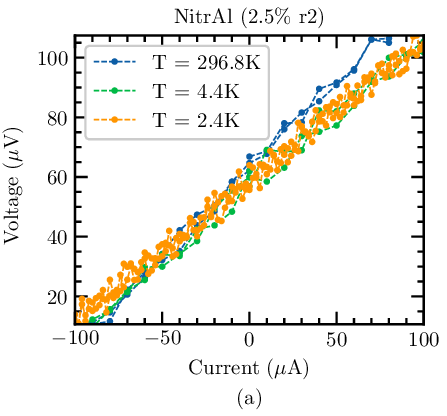}
    \quad
         \includegraphics[]{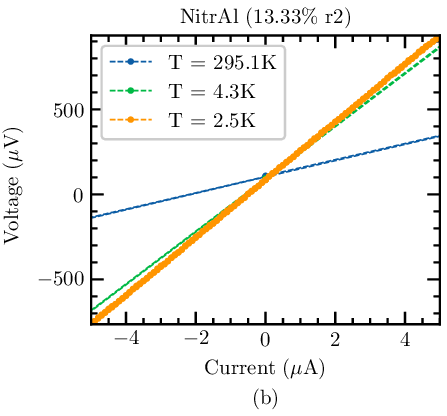}
        \caption{Current-Voltage (IV) characteristics for sample with 2.5\% \ch{N2} (a), and 13.3\% \ch{N2} (b). Notice the difference in voltage ranges used for both samples.}
        \label{fig:IVs}
\end{figure}

Figure \ref{fig:IVs} shows different IV current characteristics for different NitrAl samples. The figure is a zoom into the range used for the low temperature $T_{\mathrm{c}}$ measurements. There is a recurring offset in the voltage axis related to the source-meter setup.

\subsection{Resistance curves}
\subsubsection{Normalized resistance curves}
\begin{figure}[!htb]
    \centering
    \includegraphics{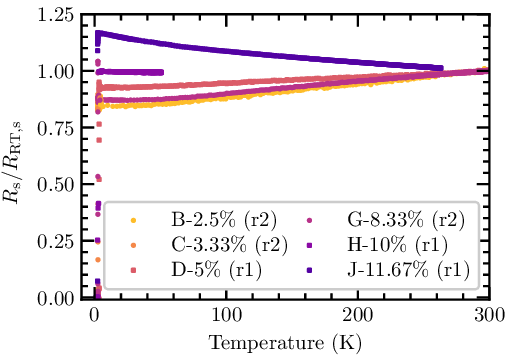}
    \caption{Normalized sheet resistance curves for several NitrAl samples.}
    \label{fig:RvsT_annex}
\end{figure}

In Figure \ref{fig:RvsT_annex} we provide the normalized sheet resistance curves not shown in the main text.

\subsubsection{Resistance peak near the superconducting transition}
\label{sec:ann_b_2}
\begin{figure}[!htb]
    \centering
    \includegraphics[width=0.3\textwidth]{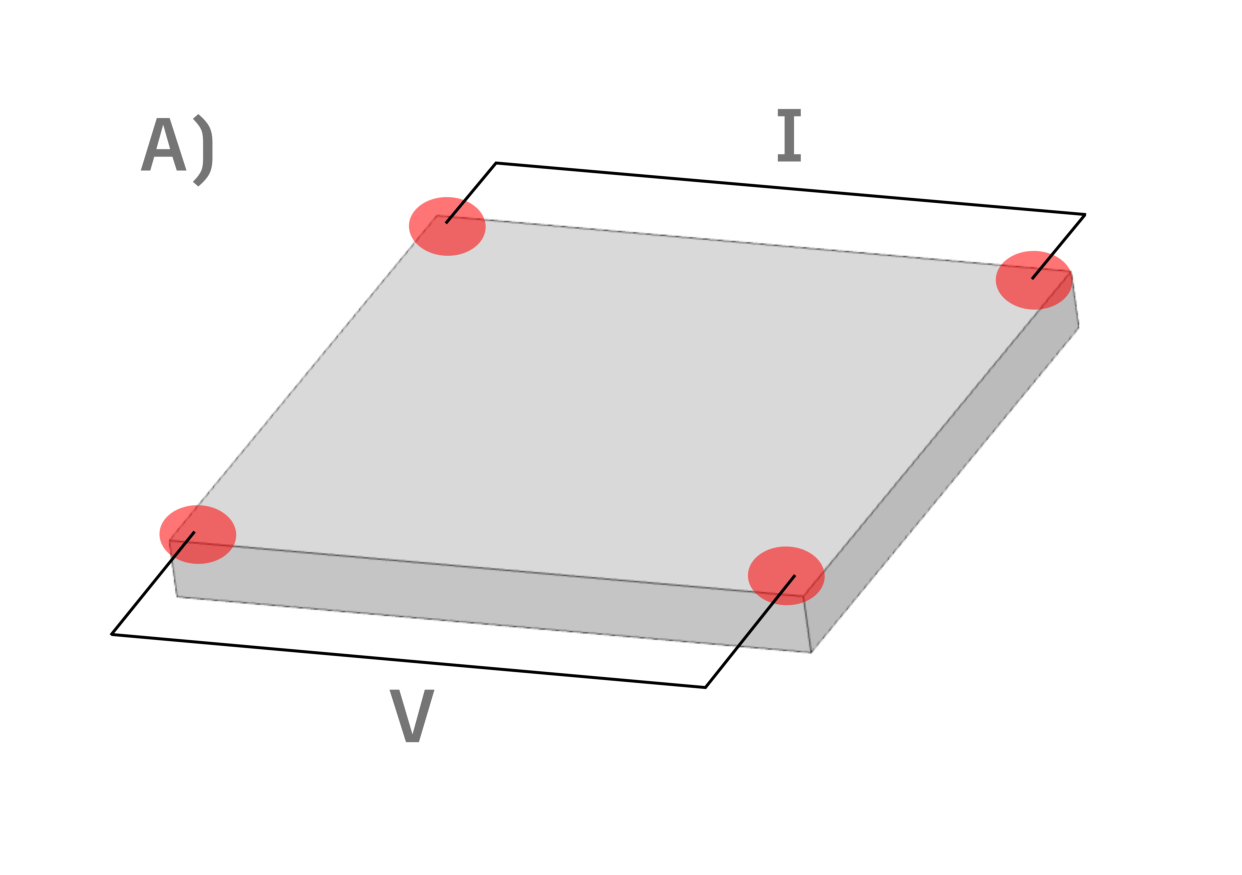}
    \quad
    \includegraphics[width=0.3\textwidth]{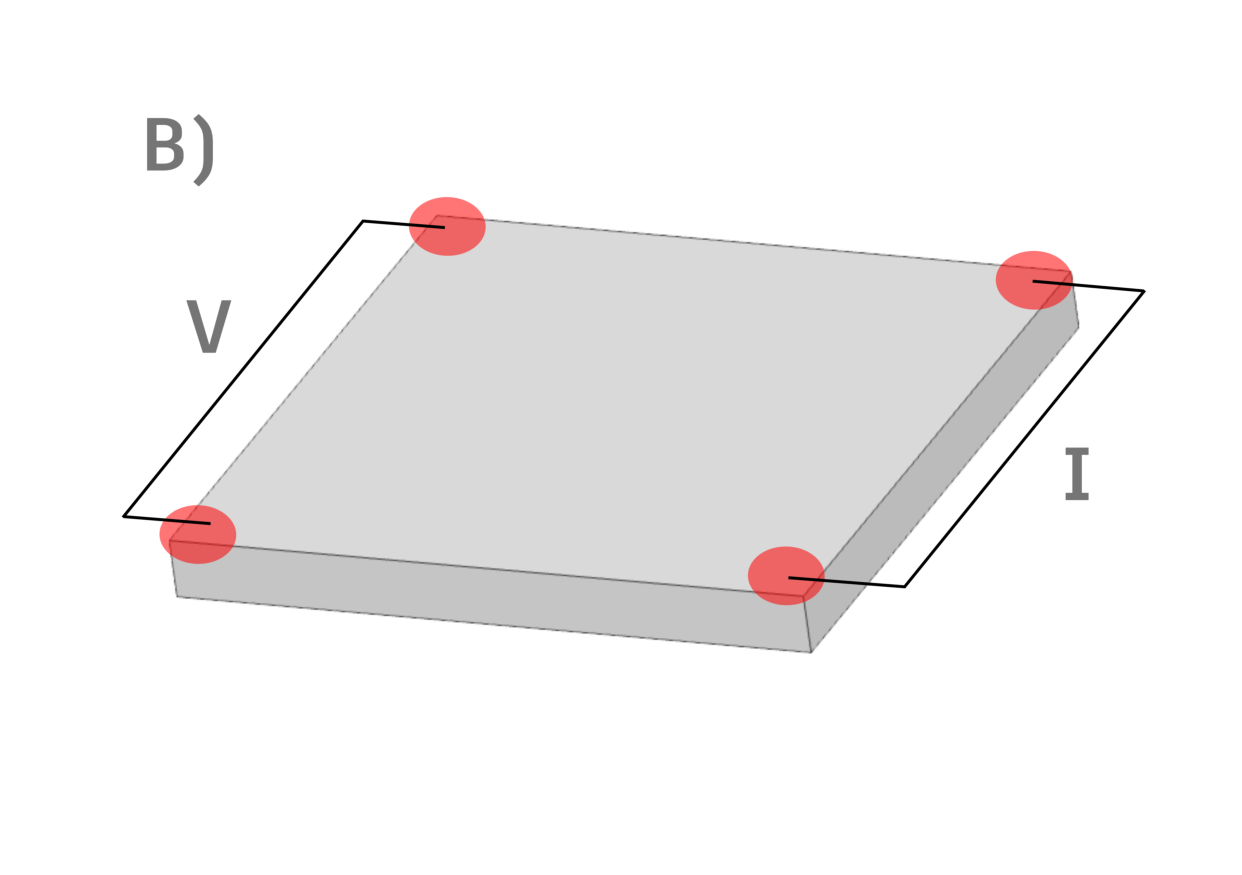}
    \quad
    \includegraphics[width=0.3\textwidth]{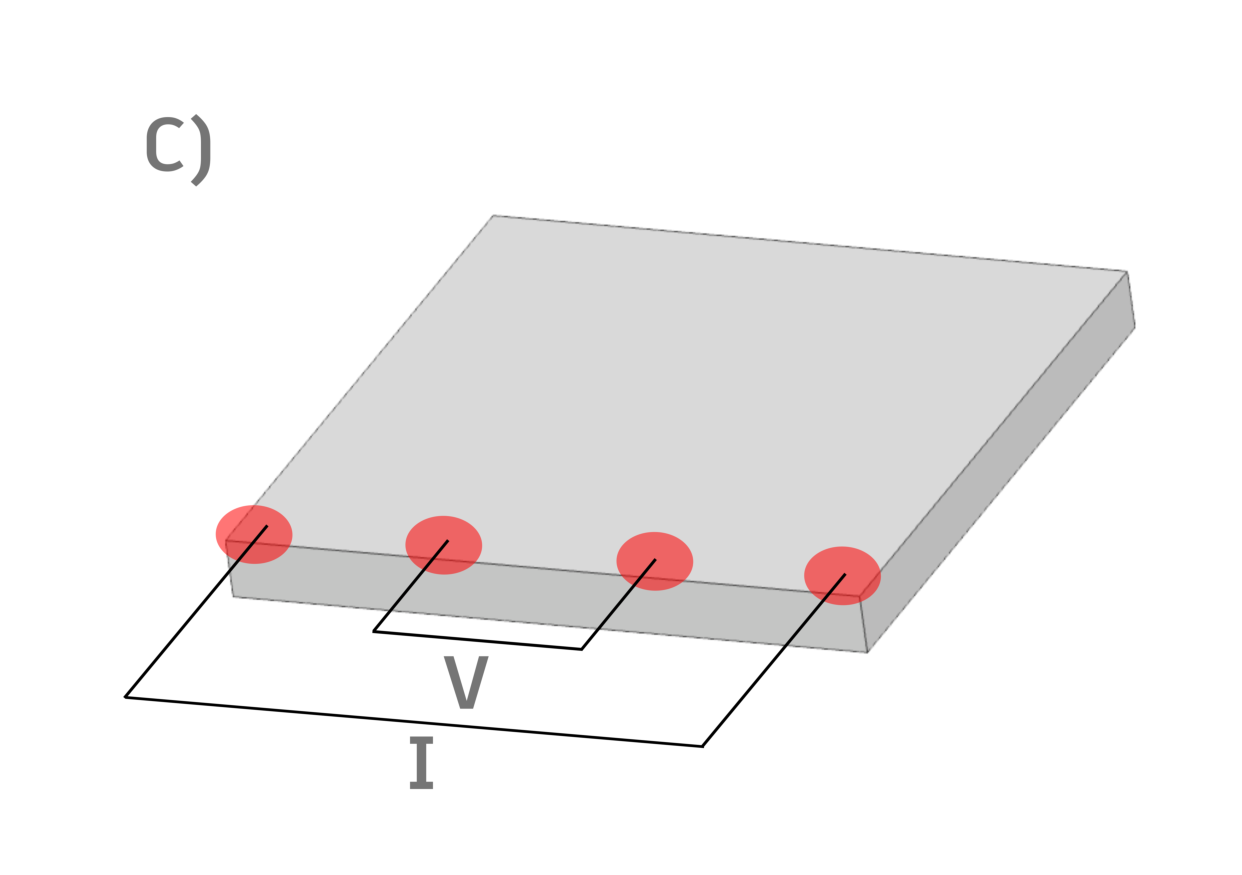}
    \caption{Different contact locations for the four probe configuration used to test the origin of the anomalous peak near the superconducting transition. }
    \label{fig:schematics_4probe}
\end{figure}

\begin{figure}[!htb]
    \centering
    \includegraphics{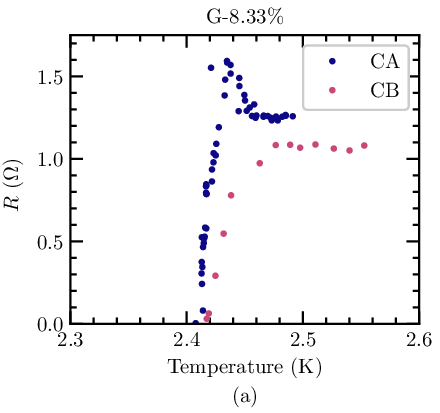}
    \quad
    \includegraphics{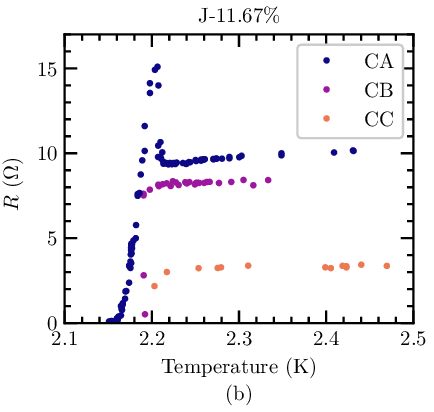}
    \caption{Resistance versus temperature for NitrAl (a) 8.33\% in two different 4-probe configurations and (b) 11.67\% in three different 4-probe configurations. }
    \label{fig:peak_study}
\end{figure}

In several samples we have observed an increase of 20\%-30\% of the resistance prior to the transition to the superconducting state. Such a peak appears only for the combination of current and voltage probes giving slightly higher resistance. We tested different probe configurations (see Figure \ref{fig:schematics_4probe}) for two samples near the superconducting transition. In Figure \ref{fig:peak_study} we plot the resistance versus temperature for different probe configurations labeled as CA, CB, CC following the nomenclature of Figure \ref{fig:schematics_4probe}. We did not include the in-line configuration C for sample G-8.33\% due to the low resistance and noise in the measurement.

With the first configuration (A) we measured slightly larger resistances and we observed the spike before the superconducting transition. Rotating the contacts (B) provided slightly lower resistance and we were not able to reproduce the rapid increase in the R(T) curve near Tc. Finally, with the in-line configuration (C), no spike was visible. A similar phenomenon has been observed in other disordered and homogeneous thin films measured in 4-probe configuration \cite{polavckova2023probing, vaglio1993explanation}. The anomalous peak can be attributed to local variations of Tc and current redistribution effects on the sample. Actually, the critical temperatures measured in different 4-probe configurations differ slightly by 0.01K.

\subsection{Sheet kinetic inductance and London penetration depth}

\begin{figure}[!hbt]
         \includegraphics[]{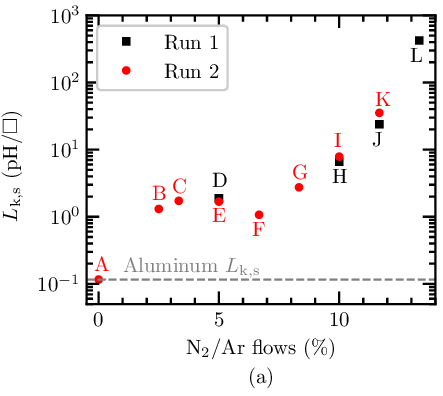}
    \quad
         \includegraphics[]{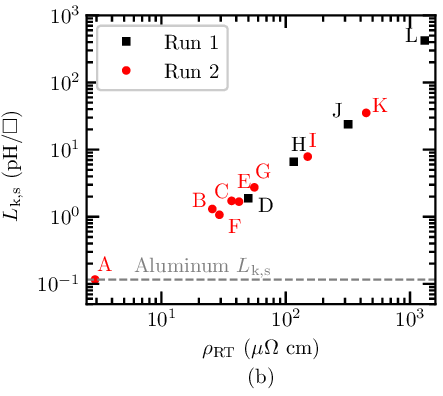}
        \caption{Estimated sheet kinetic inductance using Eq.~(\ref{eq:Kinetic_inductance}) for different \ch{N2} flow values (a), and as function of room temperature resistivity $\rho$ (b).}
        \label{fig:Lkin_plots}
\end{figure}

Using Eq.~(\ref{eq:Kinetic_inductance}), in Fig.~\ref{fig:Lkin_plots} we plot the resulting sheet kinetic inductance as function of (a) the \ch{N2} flow and (b) as function of the room temperature resistivity. 

Figure~\ref{fig:Lkin_plots}(a) shows a similar trend as $\rho$ at room temperature, with an initial jump with respect to pure Al, an intermediate regime with constant values, and a final exponential increase above 10\%.

\begin{figure}[!htb]
    \centering
    \includegraphics{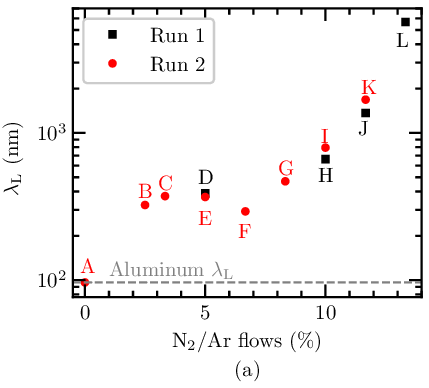}
    \quad
    \includegraphics{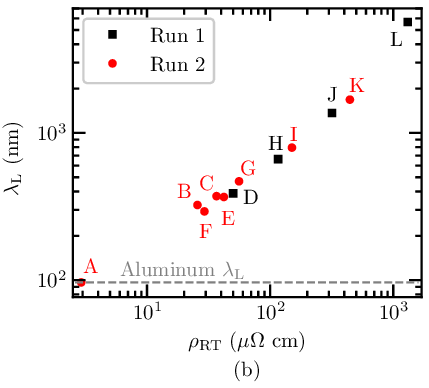}
    \caption{London penetration depth estimated for different NitrAl samples as a function of (a)  \ch{N2} flow values, and (b) room temperature resistivity $\rho$.}
    \label{fig:london_pen_depth}
\end{figure}

Similarly, in Figure \ref{fig:london_pen_depth} we plot the London penetration depth estimated using the following expression \cite{hake1967paramagnetic}, 
\begin{equation}
    \lambda_{\mathrm{L}} \simeq \SI{105}{\nano\meter}\sqrt{\frac{\rho_{\mathrm{4K}}(\SI{}{\micro\ohm\centi\meter})}{T_c (\SI{}{\kelvin})}}.
\end{equation}

\subsection{Magnetic field measurements}

\subsubsection{Experimental setup }
\label{sec:S-C1}
The samples are connected to the I-V measurement setup by wire bonding, in a simple four probe arrangement. An ADWIN gold data acquisition system and NanoQt interface was used to acquire amplified voltage data as well as apply bias voltage through a high resistor to obtain a bias current through the NitrAl thin-film samples. Such a simple configuration works as a constant current source, since the NitrAl film resistances used in these measurements at room temperature are very small ($<30~\Omega$) compared to the high-resistance bias resistor ($R_b = 10$~k$\Omega$). The maximum bias current $I_b$ was limited to 1~mA by the maximum voltage $V_b$ range of $\pm10~V$ from the ADWIN output. A Femto commercial voltage amplifier was used to with a gain 100/ 1000/ 10000 depending on the magnitude of the small voltage at the sample.

In addition to the direct I-V measurements, dV/dI at each bias current were measured by an integrated lock-in code in the ADWIN interfaced with NanoQt. A small oscillating bias current modulation of amplitude $\SI{10}{\micro\ampere}$ and frequency 91~Hz was added about each $I_b$ to measure sample resistance (dV/dI) in the lock-in configuration. The in-phase and 90 degree phase shifted components of this oscillating signal were used as a reference to measure the in-phase and out of phase components of the response voltage, which were filtered by a fourth order filter with time constant set to 50~ms (about 5 oscillations at signal frequency). The squared sum of these in-phase and out of phase components provides $R^2$. The bias current $I_b$ was varied very slowly such that during a lock-in data acquisition, the $I_b$ change can be ignored. The dV/dI obtained from lock-in measurement and from the slope of direct I-V data were compared to find that both methods provide similar results. However, the lock-in measurements show less noise, hence the resistance measured only in lock-in configuration are presented. 

\subsubsection{IV-characteristics under magnetic field}
As a complement to the data shown in Sec.~\ref{sec:4} of the main text, we show here raw data of the IV-characteristics. Note that the maximum current of the source employed is 1~mA. 

\begin{figure*}[!hbt]
    \includegraphics[width=0.92\textwidth]{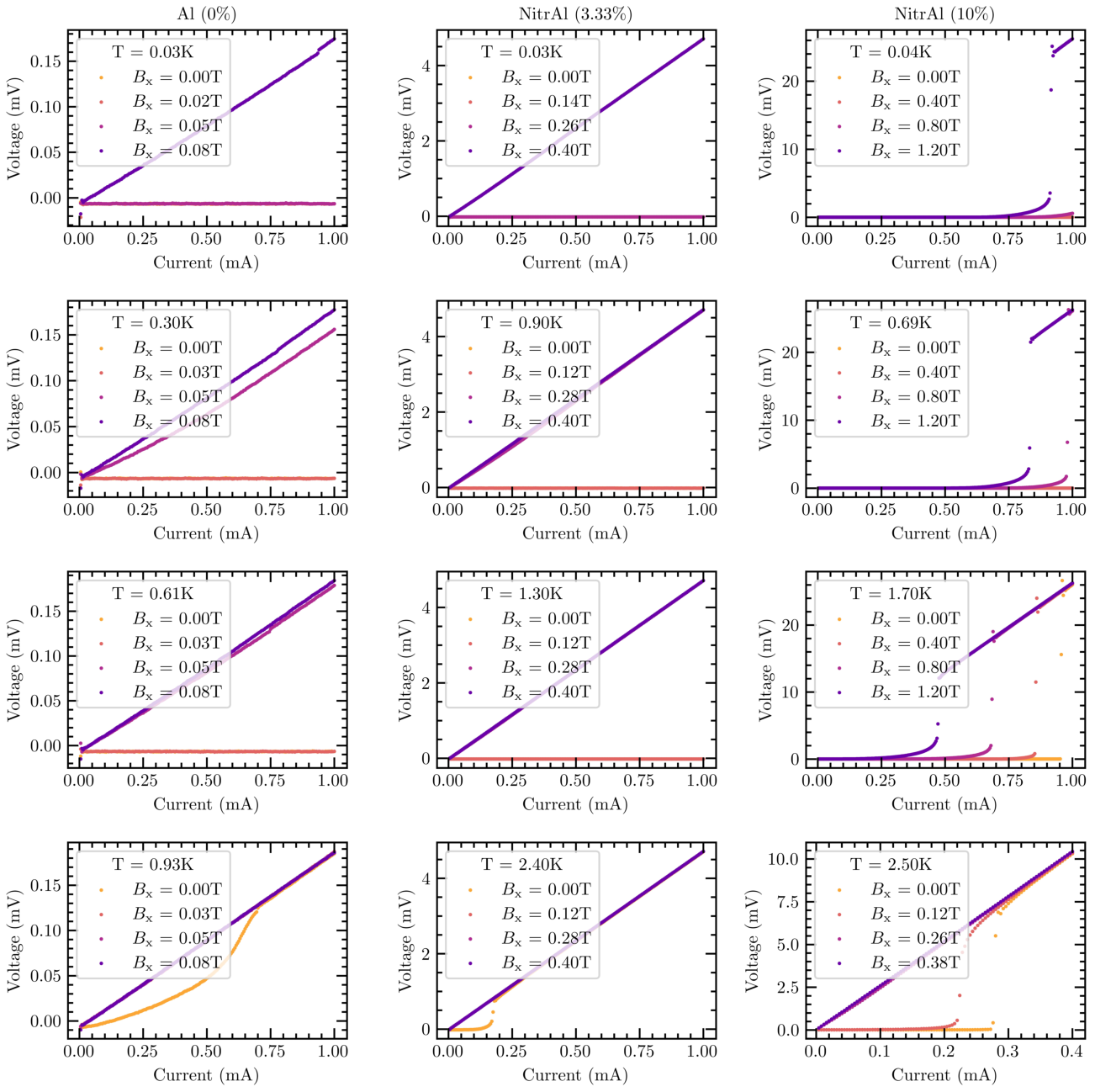}
    \caption{VI-characteristics as a function of the temperature and the magnetic field in the $x$-direction, $B_{\mathrm{x}}$ (in-plane with the chip) for different NitrAl samples produced in the second run of fabrication.}
    \label{fig:IVs_B}
\end{figure*}

\newpage
\subsection{Optical properties}
\label{sec:App-D}
\begin{figure}[!htb]
    \centering
    \includegraphics[width = 0.5\textwidth]{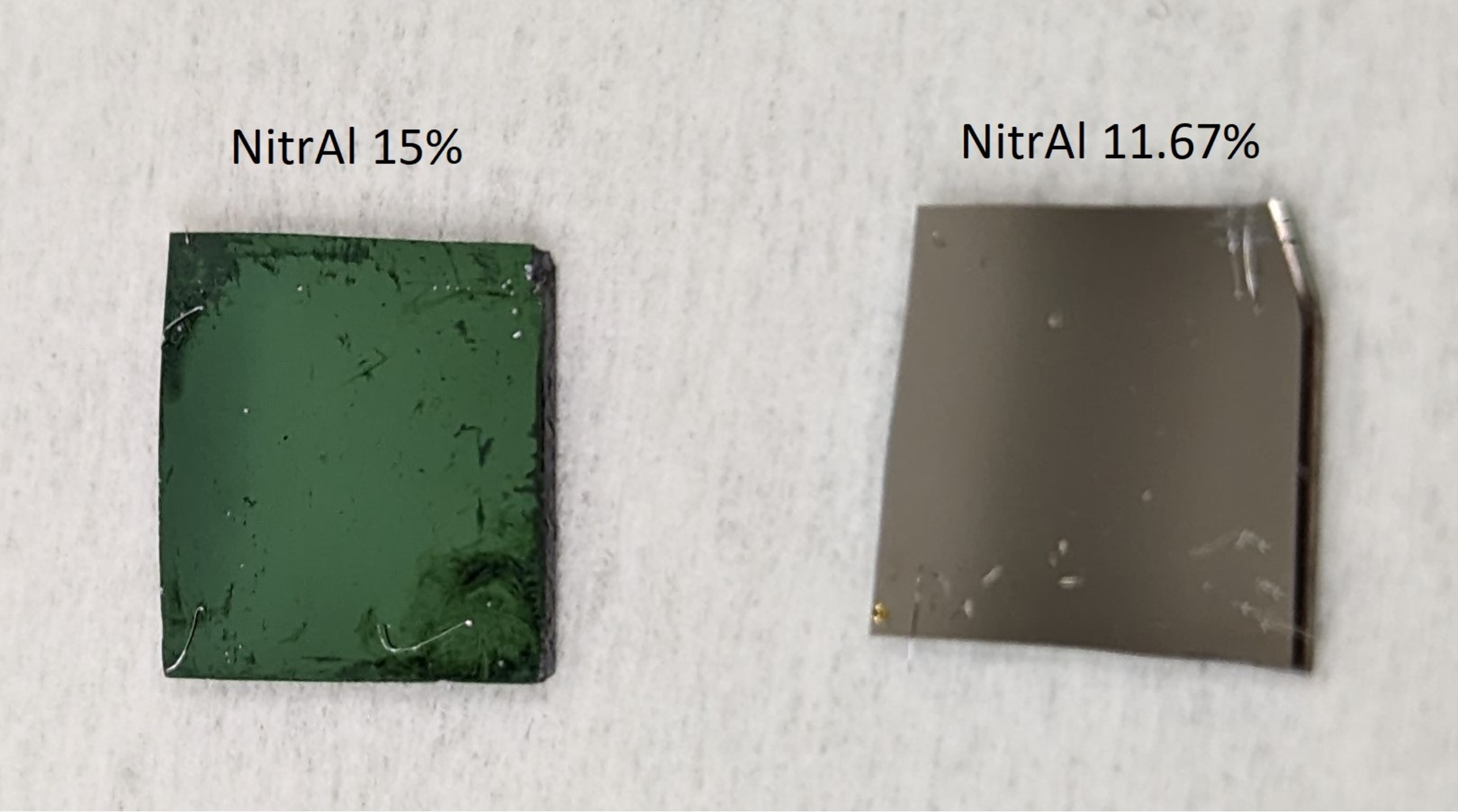}
    \caption{NitrAl films produced with different $\ch{N2}/\ch{Ar}$ flows. A clear difference in color between the insulating sample ($15\%$), of dark green, and the conductive one ($11.67\%$), of dark gray, can be observed. }
    \label{fig:color_films}
\end{figure}

We can also connect the resistivity at room temperature with the surface appearance of the samples (Fig.~\ref{fig:color_films}). A qualitative difference in the surface color of the films is observed between conductive and insulating samples. These differences might indicate a change in the optical properties and optical responses of the films. In particular, for conducting samples the surface is shiny and silver-looking, close to the one of aluminum, while samples above $13.3\%$ appear to be silver-green. A quantitative analysis of the optical response of the thin films at room temperature will be included in a future work. 

\end{document}